\documentclass{article}

\PassOptionsToPackage{numbers}{natbib}\usepackage[dblblindworkshop,final]{neurips_2025}  

\workshoptitle{The Second Workshop on GenAI for Health: Potential, Trust, and Policy Compliance}

\makeatletter
\providecommand{\@trackname}{}
\makeatother

\usepackage[utf8]{inputenc}
\usepackage[T1]{fontenc}
\usepackage{url}
\usepackage{booktabs}
\usepackage{amsfonts}
\usepackage{nicefrac}
\usepackage{microtype}
\usepackage{xcolor}
\usepackage{graphicx}
\usepackage{amsmath}
\usepackage{longtable}
\usepackage{tabularx}
\usepackage{makecell}
\usepackage{placeins}
\usepackage{listings}
\lstdefinestyle{promptbox}{
    basicstyle=\ttfamily\small,
    backgroundcolor=\color{gray!10},
    frame=single,
    breaklines=true,
    breakatwhitespace=true,
    tabsize=2,
    showstringspaces=false
}

\newcolumntype{L}{>{\raggedright\arraybackslash}X}
\newcolumntype{C}{>{\centering\arraybackslash}X}

\title{Mind the Gap: Aligning Knowledge Bases with User Needs to Enhance Mental Health Retrieval}

\author{%
  Amanda Chan \\
  Princeton University\\ USA \\
  \And
  James Liu \\
  National University of Singapore \\
  Singapore \\
  \AND
  He Kai \\
  National University of Singapore \\
  Singapore \\
  \And
  Onno P. Kampman \\
  MOH Office for Healthcare Transformation \\
  Singapore \\
}

\begin{document}

\maketitle

\begin{abstract}
    Access to reliable mental health information is vital for early help-seeking, yet expanding knowledge bases is resource-intensive and often misaligned with user needs.
    This results in poor performance of retrieval systems when presented concerns are not covered or expressed in informal or contextualized language.
    We present an AI-based gap-informed framework for corpus augmentation that authentically identifies underrepresented topics (gaps) by overlaying naturalistic user data such as forum posts in order to prioritize expansions based on coverage and usefulness.
    In a case study, we compare Directed (gap-informed augmentations) with Non-Directed augmentation (random additions), evaluating the relevance and usefulness of retrieved information across four retrieval-augmented generation (RAG) pipelines.
    Directed augmentation achieved near-optimal performance with modest expansions--requiring only a 42\% increase for Query Transformation, 74\% for Reranking and Hierarchical, and 318\% for Baseline--to reach $\sim$95\% of the performance of an exhaustive reference corpus.
    In contrast, Non-Directed augmentation required substantially larger and thus practically infeasible expansions to achieve comparable performance (232\%, 318\%, 403\%, and 763\%, respectively).
    These results show that strategically targeted corpus growth can reduce content creation demands while sustaining high retrieval and provision quality, offering a scalable approach for building trusted health information repositories and supporting generative AI applications in high-stakes domains.
\end{abstract}


\section{Introduction}

Access to reliable and timely mental health information is essential for early help-seeking and psychological support~\citep{pretorius2019young}.
\citet{kalckreuth2014mental} found that 70.9\% of psychiatric patients used the Internet for mental health purposes, most often to learn about disorders, medications, or services.
With the growing reliance on self-help platforms, forums, chatbots, and apps~\citep{he2019understanding}, retrieving accurate and trusted psychoeducational resources has become increasingly important--both when accessed directly or when surfaced in chatbot responses through retrieval-augmented generation (RAG)~\citep{kampman2024multi}.

RAG is a natural language processing approach that bridges fluent generation and factual grounding by combining the semantic strengths of large language models (LLMs) with evidence from trusted knowledge bases~\citep{oche2025systematic, lewis2020retrieval}.
By conditioning outputs on retrieved documents rather than relying solely on pretrained parameters, RAG reduces hallucinations and produces more accurate, contextually appropriate, and trustworthy responses~\citep{ayala-bechard-2024-reducing, chung2023challenges, sharma2025retrieval, he2025external}.

In mental health, knowledge bases that power retrieval systems usually contain articles, exercises, and therapeutic guides curated by trained professionals.
These are often national platforms tailored to local contexts (e.g., seasonal affective disorder is absent in tropical countries like Singapore), but they are not designed for automated retrieval and are costly to expand--especially with evolving user needs.
As a result, they often lack coverage across the full spectrum of user needs, both in content and in alignment with user style, language, and context~\citep{ibrahim2021automated, mao2022biases, he2022meta}.
For instance, a systematic review inventory by the Swedish Agency for Health Technology Assessment (SBU) identified over 2,000 evidence gaps in mental health from 2005–2020, highlighting severe content gaps where no systematic review exists or where evidence remains inconclusive~\citep{sbu2021evidencegaps}.
We posit that one powerful way of characterizing such gaps is to overlay the knowledge base with an organic source of expression of user needs: gaps occur when topics are frequently raised by users in queries, surveys, or helpline conversations but are underrepresented or absent in the corpus.
Ultimately, the effectiveness of information retrieval depends on the quality and completeness of the underlying corpus~\citep{li2024you, he2017enriching}.

Recent studies have examined how gaps between user demand and document supply affect retrieval and content generation.
\citet{abian2022wikidata} showed that such misalignments can be systematically identified, but their coverage analysis relies on Wikipedia pageview metrics and lacks generalizability beyond that domain.
\citet{sinha2023findability} introduced the metric of \textit{findability}, which measures how easily relevant documents can be surfaced by real queries, distinguishing it from traditional retrievability metrics.
However, findability remains a passive diagnostic that does not actively resolve coverage gaps.
In contrast, \citet{kang2025improving} proposed CCQGen, a concept-coverage–based query generation framework that iteratively produces synthetic queries to expose uncovered concepts within individual documents.
While CCQGen improves internal document representation, it does not incorporate real user demand or address under-served topics at the corpus level.

More recent work has explored related approaches that associate each document with synthetic queries to enhance retrieval.
For example, \citet{raina2024question} generate question-answer pairs over atomic document units to better align user queries with content, while \citet{yang2024geometry} apply a similar idea in healthcare by linking each document to answerable synthetic queries.
These methods improve retrieval within existing corpora, whereas our work focuses on corpus-level augmentation.
Specifically, we integrate demand-driven gap detection with adaptive synthetic content generation to identify under-served topics and create targeted new content to close these gaps.

In this paper, we take a systematic and automated approach to improving RAG methods at the source: finding and filling content gaps rather than only optimizing retrieval or ranking.
To demonstrate the utility of our framework, we apply it to a large mental health knowledge base.
Our contributions are threefold:

\begin{itemize}
    \item \textbf{Knowledge base gap analysis using real-world conversations}. We present a systematic framework for identifying topic-level content gaps. Using the mental health resources from \textit{mindline.sg}~\citep{weng2024mental}, Singapore’s national digital mental health platform, as a case study, we show how real users’ posts on the \textit{let’s talk} anonymous mental health forum can reveal underserved needs.
    \item \textbf{Synthetic document augmentation}. We demonstrate how these gaps can inform efficient knowledge base augmentation by recommending the content of documents to be added. We show that strategically enriching the corpus with synthetic documents targeted at critical gaps improves retrieval performance, with directed augmentation achieving near–large-scale performance at a fraction of the size.
    \item \textbf{Empirical evaluation}. We evaluate the relevance and usefulness of recommended resources using LLM-as-a-Judge techniques. Our results are robust across multiple RAG pipelines, including baseline, hierarchical, reranking, and query transformation methods.
\end{itemize}

Our findings provide practical guidance for developers of digital mental health tools and content platforms.
More broadly, this work demonstrates how retrieval systems can be systematically aligned with user needs.

\section{Methodology}

\begin{figure}
  \includegraphics[width=0.9\textwidth]{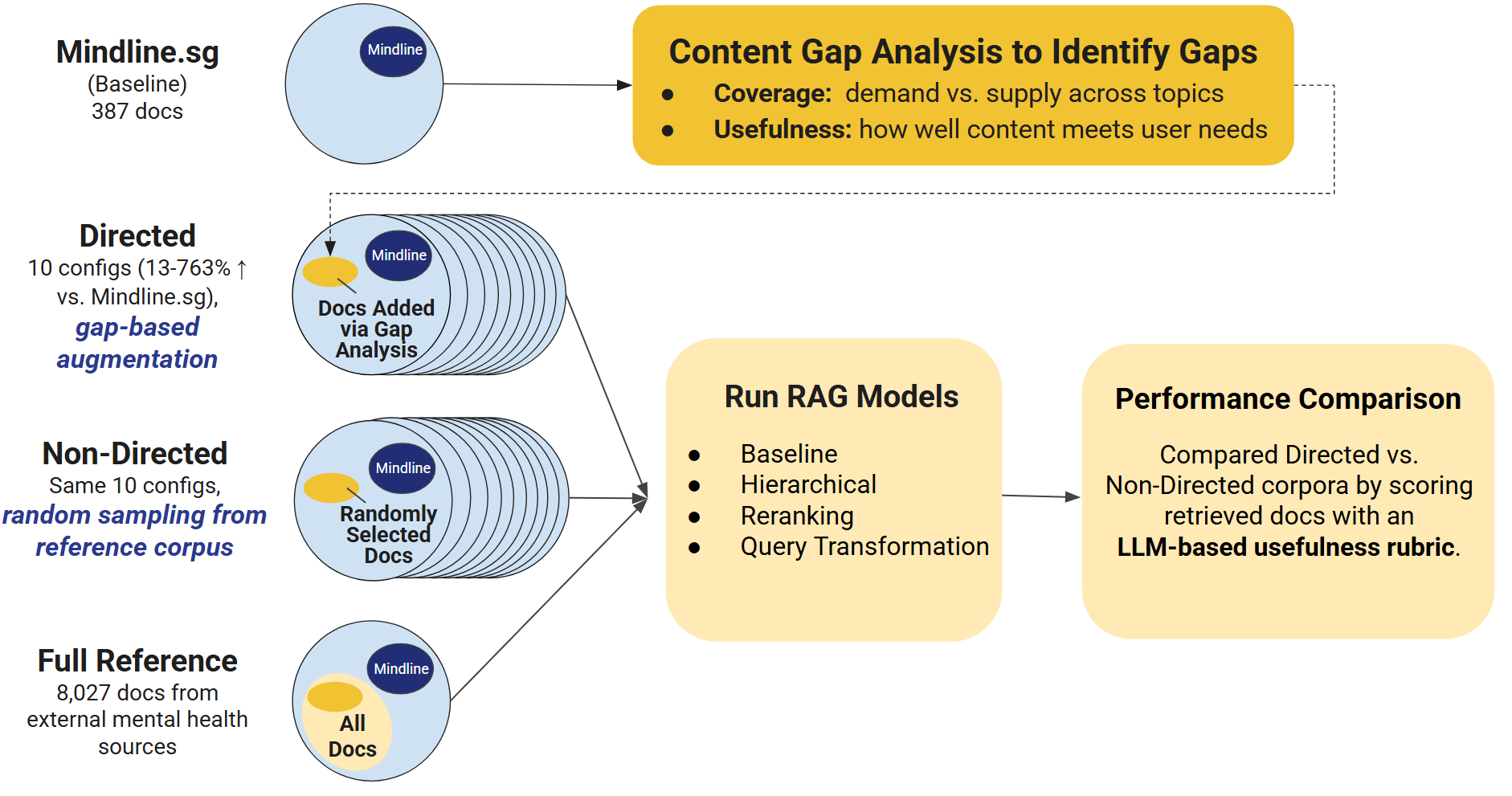}
  \caption{Overview of gap-informed corpus augmentation and evaluation framework. Forum posts are used to assess gaps in a mental health resource library, which is then augmented with additional articles and evaluated through multiple retrieval techniques.}
  \label{fig:finaloverview}
\end{figure}

Our framework consists of four stages: (1) data collection and preprocessing, (2) content gap analysis, (3) synthetic document generation and corpus augmentation, and (4) automated evaluation of retrieval quality across multiple RAG pipelines (Figure~\ref{fig:finaloverview}).
In the first stage, we assembled two types of data: a knowledge base of mental health resources and a set of real user queries expressing information needs.
Next, we analyzed these sources together to identify content gaps, areas where user needs were not well matched by available resources.
We then expanded the knowledge base with additional documents by generating synthetic content inspired by external sources.
Finally, we evaluated how well different versions of the knowledge base supported retrieval across four RAG pipelines, allowing us to compare targeted, gap-informed expansions with random expansions and test whether filling content gaps can achieve high retrieval quality with fewer documents overall.

\subsection{Data Collection and Preprocessing}

We used two primary data sources: a mental health corpus of psychoeducational resources and a real-world source of expressed user concerns.
The knowledge base consisted of 387 curated documents from \textit{\textit{\textit{mindline.sg}}}~\citep{weng2024mental}.
The naturalistically expressed user queries were based on 1,223 publicly available, anonymous posts in the ''Ask a Therapist'' subforum of \textit{let's talk} where users can pose questions for professional counselors, thus indicating individual needs~\citep{letstalk2025}.
The posts were split into 978 (80\%) training queries for gap identification and 245 (20\%) test queries for evaluation.

\subsubsection{Topic Taxonomy and Annotation}

Resources and user queries were categorized into the same list of topics to enable matching. We based our topic taxonomy on the \textit{Clinician Index of Client Concerns} (CLICC), a framework used by mental health professionals to categorize concerns presented by clients~\citep{center2019center}.
To improve interpretability of these 46 CLICC topics, we expanded each topic into eight more granular subtopics, refined with therapist input, yielding 368 clinically relevant categories (see Appendix~\ref{appendix:topic_list}). Using \texttt{GPT-4o} (see Appendix~\ref{appendix:topic_prompt} for prompt), we labeled the 978 user queries and 387 \textit{\textit{\textit{mindline.sg}}} resources with a single representative topic and subtopic.

\subsection{Content Gap Analysis}

Our proposal is that knowledge base gaps can be identified by overlaying naturalistic data indicating \textit{demand} for certain content and topics.
We demonstrate this with posts from a mental health forum.
We applied a gap analysis framework to the \textit{\textit{\textit{mindline.sg}}} corpus, assigning each query and document a subtopic for topic-level scoring.
We defined two metrics, \textit{coverage gap} and \textit{usefulness gap}, which were computed and combined into an equal-weight hybrid gap.

\subsubsection{Coverage Gap: TF-IDF–Inspired Scoring}

We designed a metric to capture coverage of topics, measuring how well a subtopic’s real-world demand matches available supply, without analyzing the content of queries or resources beyond the subtopic label.
\textit{Demand} is how often users raise a subtopic in queries, and \textit{supply} is the number of documents in the corpus covering it.
A high score signals a coverage gap, where a topic is frequently discussed or asked about but under-resourced.  

This metric is inspired by the Term Frequency-Inverse Document Frequency (TF-IDF) formula~\citep{tfidf}, which measures how important a word is to a particular document in a collection by weighing a term’s local frequency against its global rarity:

\begin{equation}
    \text{TF-IDF}(t, d) = \frac{f_{t,d}}{\max_k f_{k,d}} \cdot \log \left( \frac{N}{df(t)} \right),
\end{equation}

where $f_{t,d}$ is the frequency of term $t$ in document $d$, $df(t)$ is the number of documents containing $t$, and $N$ is the total number of documents.

We adapted TF--IDF from the \textit{word--document} to the \textit{subtopic--query} setting by redefining its components. 
Here, term frequency ($TF$) measures how often a \textit{subtopic} appears across user \textit{queries} (local demand), while inverse document frequency ($IDF$) measures how many \textit{documents} in the knowledge base cover that subtopic (global supply). 
Subtopics with high $TF$ but low $IDF$ indicate areas where user demand exceeds available content, quantifying coverage gaps between what users seek and what information exists.


\begin{equation}
    \text{Gap}(t) = \frac{\log(1 + f_p(t))}{\max_w \log(1 + f_p(w))} \cdot \left[ \log \left( \frac{D + c}{df(t) + c} \right) \right]^\alpha,
\end{equation}

where $f_p(t)$ is the number of queries labeled with subtopic $t$, $df(t)$ is the number of \textit{\textit{\textit{mindline.sg}}} documents covering $t$, $D$ is the total number of documents, $c = 1$ is a smoothing constant, and $\alpha = 1.5$ amplifies the effect of underrepresented topics. Thus, the score highlights subtopics with high demand but low supply, identifying knowledge gaps that limit the users' ability to find  resources.

\subsubsection{Usefulness Gap: LLM-as-a-Judge}

Our usefulness gap metric measures how well content supports user needs.
Unlike coverage gaps, which flag underrepresented topics, usefulness gaps capture cases where content exists but is vague, overly technical, or contextually misaligned.
In short, they assess whether available information is genuinely helpful.  

We adopt the LLM-as-a-Judge paradigm, where LLMs evaluate content quality.
Prior work shows that models like GPT-4 can reach about 80\% agreement with human preferences, a level similar to agreement between humans, when tested on both expert-designed questions and large-scale crowdsourced comparisons~\citep{zheng2023judgingllmasajudgemtbenchchatbot}.
This demonstrates the scalability of LLM-as-a-Judge, though limitations such as bias and reasoning gaps remain.
Recent frameworks extend this approach to RAG, focusing on \textit{Context Relevance}, \textit{Answer Faithfulness}, and \textit{Answer Relevance}~\citep{es-etal-2024-ragas, saad-falcon-etal-2024-ares} as key components of these LLM-based prompt rubrics.

We used \texttt{GPT-4o-mini}~\citep{openai2024gpt4omini} with a \emph{Therapist-Guided Usefulness Rubric} (Appendix~\ref{appendix:rubric}) to score query–document pairs sharing the same subtopic on:  

\begin{enumerate}
    \item \textbf{Contextual Relevance (1--50):} alignment with user needs, tone, specificity, and context.  
    \item \textbf{Practical Helpfulness \& Engagement (1--50):} clarity, feasibility, and actionable guidance.  
\end{enumerate}

For each query, the three highest-scoring documents were selected and averaged to yield a per-query usefulness score (0–100).
Subtopic-level scores were then computed by averaging across queries, applying min–max scaling to normalize values, and inverting them by subtracting each scaled score from 100, so that higher values correspond to larger usefulness gaps.

\subsubsection{Hybrid Metric}

We blended coverage and usefulness gap scores into a single hybrid metric to holistically assess corpus needs.
The hybrid score was weighted 50\% coverage and 50\% usefulness (or 100\% coverage when usefulness data was unavailable, i.e., when a subtopic had zero documents). 

We also conducted a sensitivity analysis varying coverage/usefulness weighting from fully coverage-based (100/0) to fully usefulness-based (0/100).
Corpus composition of targeted documents differed only up to 24.2\% (see Appendix~\ref{appendix:coverage_usefulness_weights}).
Given this stability, we adopted a balance 50/50 weighting.

\subsection{Synthetic Document Generation and Corpus Augmentation}

To evaluate our gap identification framework and its impact on retrieval quality, we augmented the \textit{mindline.sg} knowledge base with synthetic resources.
In our setting, synthetic documents were generated by language models to approximate the structure and tone of real-world psychoeducational resources.
This approach allowed us to expand coverage in areas where authentic materials were sparse.
Recent work, such as the ARES framework~\citep{saad-falcon-etal-2024-ares}, has even relied entirely on synthetic data for RAG evaluation.
In contrast, we use synthetic documents to complement the licensed and curated materials in \textit{mindline.sg}, enabling substantial expansion of the corpus and the construction of alternative corpora for testing and comparison.



We began by aggregating global mental health resources to create an exhaustive reference corpus for augmentation.
Publicly available metadata (titles, headers, word counts) were collected from seven leading platforms across the US, UK, Canada, and Australia: \textit{Verywell Mind} (3,109 docs), \textit{PositivePsychology.com} (974), \textit{PsychCentral} (421), \textit{GoodTherapy} (2,041), \textit{HeadsUpGuys} (113), \textit{ReachOut} (766), and the UK’s \textit{NHS} (216).
Due to scraping restrictions, only NHS articles were retained in full, as their public, non-personal content is licensed under the Open Government Licence.
For the other sources, we used publicly available metadata only.

Using this metadata, we generated 7,424 synthetic articles with \texttt{GPT-4o-mini}, guided by prompts emphasizing therapist-informed principles such as emotional safety, harm reduction, and inclusivity (see Appendix~\ref{appendix:synthetic_prompts}).
Each article was created in 16.1 seconds on average at a cost of \$0.0007.

Combined with the 216 licensed NHS articles, this yielded 7,640  documents from non-\textit{\textit{\textit{mindline.sg}}} platforms, which represents our reference corpus used for augmenting the original knowledge base.

\subsubsection{Construction of Directed and Non-Directed Corpora}

While knowledge bases should in theory improve with every added article, it is not practically feasible to indiscriminately include all available content.
We instead use gap analysis to guide expansion.  

To evaluate the impact of added documents, we built ten Directed corpora (gap-informed) and ten Non-Directed corpora (randomly sampled), ranging from 437 to 3,341 documents (+12.9\% to +763.3\% over the 387-document \textit{\textit{mindline.sg}} baseline).
For comparison, the full Reference Corpus (8,027 documents) represents a +1,974.2\% increase.
All corpora include the \textit{mindline.sg} baseline.  

For Directed corpora, 7,640 external documents were assigned subtopics, paired with relevant \textit{let’s talk} (training set) queries, and scored (1–100) by \texttt{GPT-4o-mini} using the \emph{Therapist-Guided Usefulness Rubric}.
To determine how many documents to allocate to each subtopic, we summed the hybrid gap scores across all subtopics and then divided each individual subtopic’s score by this total.
The resulting proportion defined that subtopic’s quota, so higher-scoring subtopics received larger allocations.
Within each subtopic, documents were ranked by average score, and the top entries were selected.
Non-Directed corpora were created by random sampling from the Reference Corpus, matched in size to each Directed configuration for direct comparison.

\begin{table}[!ht]
\caption{Corpus configurations and relative size increases.}
  \centering
  \scalebox{0.9}{
    \begin{tabular}{lccc}
      \toprule
      \textbf{Corpus} & \textbf{Added \# Docs} & \textbf{Total \# Docs} & \textbf{\% Increase from \textit{\textit{mindline.sg}}} \\
      \midrule
      \textit{\textit{mindline.sg}} (baseline) & 0     & 387   & 0\% \\
      Directed / Non-Directed 1 & 50    & 437   & +12.9\% \\
      Directed / Non-Directed 2 & 162   & 549   & +41.9\% \\
      Directed / Non-Directed 3 & 288   & 675   & +74.4\% \\
      Directed / Non-Directed 4 & 500   & 887   & +129.2\% \\
      Directed / Non-Directed 5 & 898   & 1285  & +232.0\% \\
      Directed / Non-Directed 6 & 1230  & 1617  & +317.8\% \\
      Directed / Non-Directed 7 & 1560  & 1947  & +403.1\% \\
      Directed / Non-Directed 8 & 2097  & 2484  & +542.1\% \\
      Directed / Non-Directed 9 & 2561  & 2948  & +661.8\% \\
      Directed / Non-Directed 10 & 2954 & 3341  & +763.3\% \\
      Reference  & 7640  & 8027  & +1974.2\% \\
      \bottomrule
    \end{tabular}
  }
  \label{tab:corpora_summary}
\end{table}

\subsection{Automated Evaluation of Retrieval Quality}

\subsubsection{Retrieval Pipelines}

To ensure robustness across retrieval methods, we evaluated performance using four RAG pipelines: Baseline~\citep{lewis2020retrieval}, Hierarchical~\citep{zhang2024hierarchical}, Reranking~\citep{abdallah2025rankify}, and Query Transformation~\citep{ma2023query}. 

In all pipelines, documents were embedded with OpenAI’s \texttt{text-embedding-3-small} model and stored in a FAISS vector index (cosine similarity)~\citep{douze2024faiss}.
Queries were embedded with the same model, and the retriever first produced a candidate set (up to 20 documents or chunks).
Final results always consisted of the three top-ranked documents, which were passed downstream for generation.
The key differences lay in how candidates were constructed and re-ranked:

\begin{itemize}
    \item \textbf{Baseline:} A standard RAG setup where queries were directly matched against the full-text document embeddings in FAISS. The three most similar documents were returned.
    
    \item \textbf{Hierarchical:} Extended Baseline by embedding subtitle-level chunks (title + subtitle text) rather than entire documents. For each query, the top 20 chunks were retrieved from FAISS, then merged back into their full source documents to avoid fragmentary retrieval. These candidates were re-ranked with the \emph{Therapist-Guided Usefulness Rubric}, scoring contextual relevance and practical helpfulness, and the top three were selected.
    
    \item \textbf{Reranking:} Used standard document-level FAISS retrieval to identify the top 20 candidates. These were then re-scored with GPT-4o-mini, which applied the \emph{Therapist-Guided Usefulness Rubric}. The three highest-scoring documents were retained, combining efficient dense retrieval with higher-precision LLM judgment.
    
    \item \textbf{Query Transformation:} Augmented retrieval by first rewriting each query for greater specificity and recall using a prompt from the \textit{RAG\_Techniques} repository~\citep{nir_diamant_rag_techniques}. The rewritten query was then used to retrieve 20 candidates from FAISS, which were re-ranked using the same LLM-based scoring as in the reranking pipeline. The top three were returned.
\end{itemize}

\subsubsection{Evaluation Method}

We assessed retrieval quality by running all four RAG pipelines (Baseline, Hierarchical RAG, Reranking, and Query Transformation) across 22 corpus configurations, namely \textit{mindline.sg}, ten Directed corpora, ten Non-Directed corpora, and the full Reference, totaling 88 experiments.
Evaluation used 248 held-out test queries excluded from the content gap analysis and corpus construction.
Retrieved documents were scored by \texttt{GPT-4o-mini} using the \emph{Therapist-Guided Usefulness Rubric}.
To compare pipelines and corpora, we averaged LLM usefulness scores across all 248 queries and their top three retrieved documents, yielding a single aggregate score for each experiment.
We used the top three rather than only the top document to better approximate realistic retrieval scenarios, where users are typically exposed to multiple candidate resources instead of a single result.



\section{Results}

\begin{figure}
  \centering
  \includegraphics[width=\textwidth]{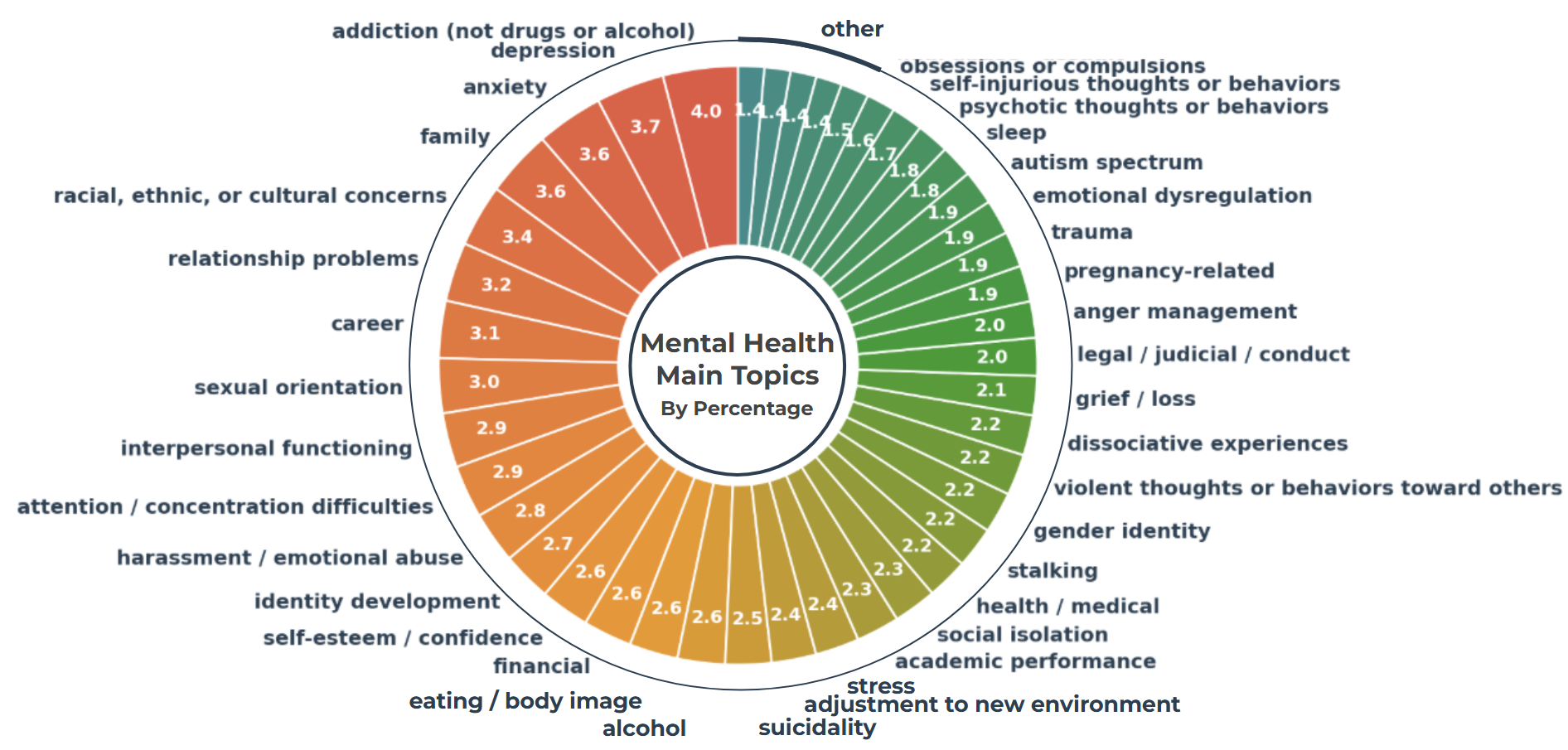}
    \caption{Proportional distribution of Directed corpus across main topics.}
    \label{fig:maintopicPHOTO}
\end{figure}

Our analysis was conducted at the \textit{subtopic level}, which allowed for finer-grained identification of specific areas of need.
For visualization purposes, however, subtopics were grouped back into their corresponding \textit{main topics} to illustrate broader thematic patterns, as shown in Figure~\ref{fig:maintopicPHOTO}.

\newpage
The top ten most under-supported topics/subtopics according to hybrid gap scores are:

\begin{enumerate}
    \item \textbf{Depression:} Self-critical thoughts and low self-worth
    \item \textbf{Relationship problems:} Trust erosion and boundary issues
    \item \textbf{Anxiety:} Fear of illness and health-related vigilance
    \item \textbf{Relationship problems:} Attachment insecurity and emotional distance
    \item \textbf{Emotional dysregulation:} Rapid mood fluctuations and reactivity
    \item \textbf{Family:} Parental conflict and household tension
    \item \textbf{Depression:} Social isolation and disconnection
    \item \textbf{Depression:} Anhedonia and withdrawal from rewarding activities
    \item \textbf{Anxiety:} Social evaluation concerns and avoidance
    \item \textbf{Anxiety:} Sleep disturbances
\end{enumerate}

For a full list of subtopics and content gap rankings, see Appendix~\ref{appendix:gap_scores}.

When aggregated at the \textit{main topic level}, the top five most under-supported categories are \textbf{Addiction (not drugs or alcohol), Depression, Anxiety, Family, and Racial, ethnic, or cultural concerns}.
Although Addiction’s subtopics do not rank in the top ten individually, they cluster just below that range (\#12--23), elevating Addiction to \#1 overall while other topics combine high- and low-ranking subtopics that lower their averages.

\begin{figure}
  \centering
  \includegraphics[width=\textwidth]{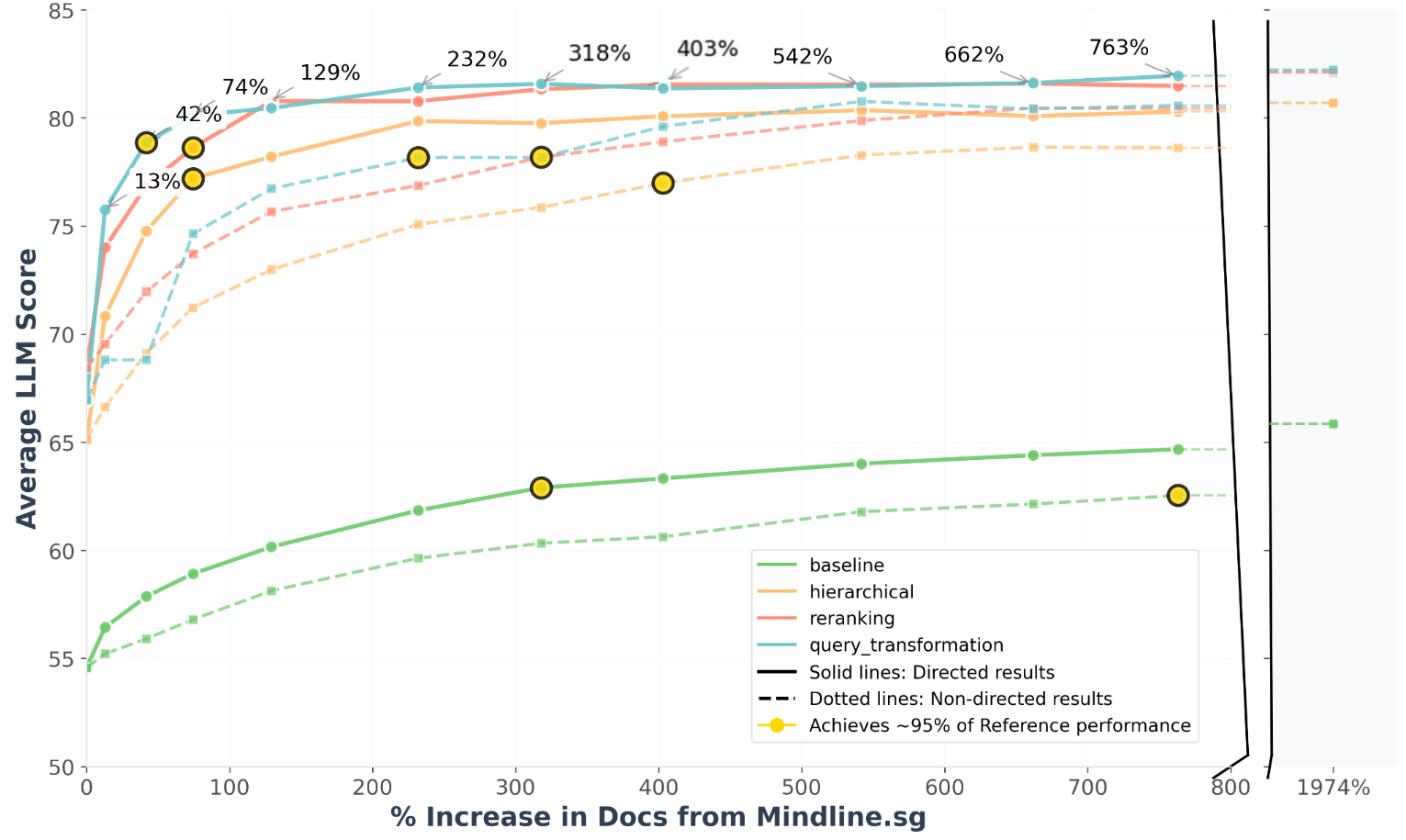}
    \caption{LLM usefulness scores by corpus size, comparing Directed vs. Non-Directed retrieval.}
    \label{fig:4trendsPHOTO}
\end{figure}

To interpret the results, instead of comparing LLM scores directly, we compare to the Reference corpus performance as a best case scenario.
RAG experiments show that gap-informed augmentation (Directed) achieved $\sim$95\% of Reference performance with far fewer documents.
The smallest Directed corpus meeting this threshold required: 42\% more docs for Query Transformation (162 docs), 74\% for Hierarchical (288 docs), 74\% for Reranking (288 docs), and 318\% for Baseline (1,230 docs) (Table~\ref{tab:doc_reduction}).
Non-Directed corpora needed more: 232\% for Query Transformation (898 docs), 318\% for Reranking (1,230 docs), 403\% for Hierarchical (1,560 docs), and 763\% for Baseline (2,954 docs).
Random sampling therefore required thousands more documents to reach the same quality.
Thus, Directed augmentation cuts content creation workload by 58.4\% (Baseline), 81.5\% (Hierarchical), 76.5\% (Reranking), and 81.9\% (Query Transformation) (Figure~\ref{fig:4trendsPHOTO}), demonstrating that gap analysis accelerates retrieval gains while minimizing corpus expansion effort.

Pipeline differences were also clear: Query Transformation performed best, reaching optimal scores with the fewest documents.
Hierarchical and Reranking were strong, while Baseline required far more resources.

Overall, these findings validate our hypothesis: even a modest, gap-informed expansion substantially improves retrieval quality.
Directed corpora rapidly approach Reference performance, while Non-Directed corpora require several times more documents.
For completeness, the full set of average LLM usefulness scores across all 88 configurations is provided in Appendix Tables ~\ref{tab:appendix_directed_rag_scores} and ~\ref{tab:appendix_nondirected_rag_scores}.

\begin{table}[t!]
  \centering
  \caption{Optimal corpus sizes to reach $\sim$95\% of Reference performance across retrieval pipelines.}
  \label{tab:doc_reduction}
  \setlength{\tabcolsep}{4pt}
  \footnotesize
  \begin{tabularx}{\linewidth}{L C C C C C}
    \toprule
    \textbf{RAG Pipeline}
      & \textbf{Directed: \% Increase in Docs from \textit{\textit{mindline.sg}}}
      & \textbf{Non-Directed: \% Increase in Docs from \textit{\textit{mindline.sg}}}
      & \textbf{Directed Docs Added}
      & \textbf{Non-Directed Docs Added}
      & \textbf{\% Decrease in Docs Added (Directed vs. Non-Directed)} \\
    \midrule
    Baseline             & 318\% & 763\% & 1230 & 2954 & 58.4\% \\
    Hierarchical         &  74\% & 403\% &  288 & 1560 & 81.5\% \\
    Reranking            &  74\% & 318\% &  288 & 1230 & 76.5\% \\
    Query Transformation &  42\% & 232\% &  162 &  898 & 81.9\% \\
    \bottomrule
  \end{tabularx}
\end{table}

\section{Discussion}

Our study shows that gap-informed corpus augmentation enables near-optimal retrieval with far fewer documents than indiscriminate expansion.
Directed corpora, guided by topic-level coverage and usefulness gaps, rapidly closed most of the performance gap to the full Reference corpus with only a few hundred added documents.
In contrast, Non-Directed corpora required thousands more to achieve comparable quality.
These findings highlight that efficient, demand-driven augmentation is not only feasible but also highly effective.

We also found differences in retrieval quality across pipelines, with Query Transformation performing best. This likely reflects the mismatch between how users informally express mental health concerns and how resources are written.
By rewriting queries into more formal, therapist-like language that aligns with document phrasing, Query Transformation improves retrieval effectiveness.

These results have meaningful real-world implications, as these knowledge bases are essential building blocks for safe and reliable AI systems.
In settings like mental health, where expert-reviewed content is costly and time-intensive to produce, a strategic approach offers a more sustainable path forward.
Rather than generating thousands of documents indiscriminately, content creators can prioritize a focused, subtopic-level document quota informed by actual user needs.
This approach is not only faster and cheaper, but also realistically helpful, providing actionable guidance for therapists and editorial teams under resource constraints.
By adapting to both content gaps and operational constraints, targeted augmentation supports scalable and efficient corpus growth without sacrificing retrieval performance.
However, we emphasize that our synthetic document generation process merely served as a proof of concept: the value proposition of most mental health knowledge bases remains that content is written by trusted experts~\citep{perivolarisOpinionMentalHealth2025}.

While gap-informed augmentation improved overall retrieval, some niche or emergent queries still performed poorly due to limited source material. Examples include relationship insecurity influenced by social media use, ongoing career or academic setbacks associated with low self-esteem, and expressions of emotional detachment or confusion about empathy. These cases indicate where human content creators should focus on new article development.

Our findings affirm that even modest, gap-informed additions can dramatically improve retrieval outcomes, closing most of the performance gap to large-scale augmentation with only a fraction of the content.
In high-stakes domains like mental health, quality beats quantity, especially when quality is guided by actual demand.
By aligning corpus growth with measured user needs, this framework offers a principled, resource-efficient strategy that can be applied across domains to deliver maximum impact with minimal expansion.
We have presented a safe application of AI for mental health care, not replacing any humans in the process, but supporting clinicians and content curators.

\section{Limitations}

While our framework highlights promising directions for content-aware mental health retrieval, several limitations should be noted. 

\textbf{Platform demographics mismatch:} \textit{let’s talk} primarily reflects youth-driven concerns, such as identity, academic stress, and relationships, whereas \textit{\textit{\textit{mindline.sg}}} targets a broader adult population with needs that include parenting, financial precarity, or grief.
Including additional sources of expressed concerns can mitigate bias in gap detection and improve generalizability across demographic groups.

\textbf{Reference corpus limitations:} Our reference knowledge base aggregates content from multiple mental health platforms, leading to some duplication.
While pruning could reduce redundancy, we note that partial overlap is typical in mental health platforms, where similar psychoeducational topics are intentionally reiterated across sites to maintain accessibility and consistency of guidance.
Future work could examine pruning or deduplication to quantify its marginal impact on retrieval quality.

\textbf{No supervised relevance labels:} All retrieval and reranking methods rely on unsupervised similarity scores or LLM-as-a-Judge evaluations without human or clinical validation.
Without ground-truth relevance annotations, the system may overlook domain-specific nuances like therapeutic framing or emotional appropriateness.
Future work should involve trained experts to evaluate the utility of this framework, allowing for aligning with clinical guidelines and review procedures before deployment.
More broadly, the lack of a standardized way to evaluate RAG in clinical settings is problematic~\citep{amugongoRetrievalAugmentedGeneration2025}.


\textbf{Synthetic-only documents:} Due to scraping restrictions on real mental health content, we used LLM-generated documents for augmentation.
While modeled after therapist-written materials, they were not medically reviewed and served only as a proof of concept.
In practice, all new content should be authored by qualified professionals to ensure clinical accuracy and maintain user trust.

\section{Conclusion}

This work demonstrates that gap-informed, directed corpus augmentation can deliver substantial efficiency gains for retrieval-augmented generation.
By systematically targeting underrepresented subtopics, our framework achieved near-optimal retrieval quality with 58--82\% fewer documents than non-directed approaches, reducing the need for costly large-scale content creation.
These findings suggest that knowledge bases in high-stakes domains such as mental health can be scaled more strategically, prioritizing coverage where it matters most while maintaining performance.
Future work could extend this approach to other domains and explore integration with dynamic feedback loops to further align corpus growth with evolving user needs.

\clearpage




\bibliographystyle{plainnat}    
\bibliography{custom} 

\clearpage


\appendix

\section{List of Main Topics and Subtopics for Categorization}
\label{appendix:topic_list}

This appendix provides the full set of main topics and verbatim subtopics used in our LLM-based categorization.
Main topics are ordered by their frequency of endorsement in the CLICC dataset, i.e., the proportion of clients for whom each concern was recorded as either a top concern or a check-all-that-apply concern~\citep{center2019center}.

\begin{longtable}{p{0.18\linewidth} p{0.78\linewidth}}
  \caption{Main topics and subtopics for categorization.} \label{tab:topics} \\

  \toprule
  \textbf{Main Topic} & \textbf{Subtopics} \\
  \midrule
  \endfirsthead

  \toprule
  \textbf{Main Topic} & \textbf{Subtopics} \\
  \midrule
  \endhead

  \midrule
  \multicolumn{2}{r}{\textit{Continued on next page}} \\
  \midrule
  \endfoot

  \bottomrule
  \endlastfoot

    Anxiety & Generalized worry and anticipatory tension; Panic symptoms and physiological reactivity; Social evaluation concerns and avoidance; Fear of illness and health-related vigilance; Uncertainty intolerance and future-focused fear; Cognitive rumination and overthinking loops; Sleep disturbances linked to anxiety; Impact of anxiety on daily functioning \\
    \midrule
    Depression & Emotional flatness and loss of affect; Anhedonia and withdrawal from rewarding activities; Cognitive distortions and negative thought patterns; Psychomotor slowing and fatigue; Feelings of hopelessness or helplessness; Self-critical thoughts and low self-worth; Social isolation and disconnection; Disrupted sleep and appetite patterns \\
    \midrule
    Stress & Acute stress responses and short-term overwhelm; Chronic stress and allostatic load; Work-life balance and professional burnout; Academic and performance-related stress; Interpersonal sources of stress; Stress-related physical symptoms; Cognitive overload and decision fatigue; Coping capacity and resilience thresholds \\
    \midrule
    Family & Parental conflict and household tension; Parent-child relationship challenges; Sibling rivalry or comparison stress; Intergenerational value conflicts; Communication breakdowns within the family; Family roles and expectations; Cultural or religious pressures in family dynamics; Navigating major family transitions or disruptions \\
    \midrule
    Relationship Problems & Romantic conflict and unresolved tension; Attachment insecurity and emotional distance; Trust erosion and boundary issues; Communication mismatches and misunderstandings; Breakups, separation, and emotional recovery; Power dynamics and control struggles; Navigating long-distance or commitment uncertainty; Impact of external stressors on relationships \\
        \midrule
    Interpersonal Functioning & Assertiveness and self-expression difficulties; People-pleasing and fear of rejection; Boundary setting and personal space; Reading and responding to social cues; Group dynamics and belongingness; Conflict resolution and emotional regulation; Dependence versus autonomy in relationships; Patterns of miscommunication or avoidance \\
    \midrule
    Self-Esteem / Confidence & Negative self-concept and inner critic; Fear of judgment and social comparison; Perfectionism and fear of failure; Self-validation versus external validation; Confidence in academic or work performance; Body image and physical self-perception; Imposter feelings and perceived inadequacy; Impact of self-esteem on decision-making \\
    \midrule
    Trauma & Hyperarousal and physiological reactivity; Intrusive memories and flashbacks; Avoidance of trauma reminders; Emotional numbing and detachment; Interpersonal trust and boundary disruptions; Shame, guilt, or self-blame post-trauma; Disrupted memory or time perception; Reenactment cycles and maladaptive coping \\
    \midrule
    Sleep & Insomnia and difficulties initiating sleep; Sleep maintenance and frequent awakenings; Nightmares and emotionally intense dreaming; Circadian rhythm misalignment; Stress-related sleep disturbances; Restorative quality of sleep; Impact of screen time or stimulation; Sleep and daytime fatigue or functioning \\
    \midrule
    Eating / Body Image & Body dissatisfaction and self-surveillance; Restrictive eating patterns and control; Emotional or binge eating behaviors; Preoccupation with weight and shape; Cultural and societal appearance standards; Body checking and comparison habits; Internalized stigma around food and eating; Fluctuating self-esteem linked to body image \\
    \midrule
    Adjustment to New Environment & Cultural adaptation and value conflicts; Loss of familiarity and support systems; Navigating social norms in new settings; Emotional responses to relocation; Uncertainty and identity shifts; Language or communication challenges; Routine disruption and instability; Isolation during transitional periods \\
    \midrule
    Social Isolation & Loneliness despite social presence; Emotional withdrawal from connection; Avoidance of group or public interaction; Disconnection from meaningful relationships; Perceived rejection or exclusion; Challenges in initiating or sustaining friendships; Fear of vulnerability in social spaces; Impact of isolation on emotional health \\
    \midrule
    Academic Performance & Procrastination and avoidance cycles; Fear of failure and perfectionistic standards; Distractibility during study or tasks; Low academic motivation or engagement; Test anxiety and performance pressure; Imposter syndrome in academic settings; Cognitive overload and burnout; Difficulty with executive functioning \\
    \midrule
    Grief / Loss & Shock and disbelief following loss; Sadness and yearning for what was lost; Disruption to roles and daily routines; Emptiness and identity confusion; Ambiguous or disenfranchised grief; Cultural or spiritual grief expressions; Social withdrawal during mourning; Difficulty accepting permanence of loss \\
        \midrule
    Suicidality & Hopelessness and desire for relief; Passive versus active suicidal thoughts; Ambivalence about living or dying; Internalized stigma and silence; Suicidal ideation during emotional distress; Barriers to help-seeking or disclosure; Impact of suicidality on relationships; History of self-harm or prior attempts \\
    \midrule
    Attention / Concentration Difficulties & Inconsistent ability to sustain attention; Distractions from internal or external stimuli; Working memory limitations and forgetfulness; Mental fatigue after prolonged focus; Difficulty switching between tasks; Procrastination linked to focus challenges; Overstimulation in busy environments; Task initiation and follow-through issues \\
    \midrule
    Sexual Abuse / Assault & Violation of bodily autonomy and trust; Fear and hyperarousal in safe spaces; Shame and internalized self-blame; Avoidance of reminders or triggers; Emotional numbing and dissociation; Difficulty with intimacy or touch; Barriers to disclosure or seeking support; Ongoing safety concerns and vigilance \\
    \midrule
    Identity Development & Exploration of roles, values, and beliefs; Uncertainty in self-definition or direction; Social comparison and identity confusion; Cultural or gender identity exploration; Balancing individuality and group belonging; Internal conflict around authenticity; Shifts in identity across transitions; Intersectionality and layered identities \\
    \midrule
    Emotional Dysregulation & Rapid mood fluctuations and reactivity; Difficulty identifying or naming emotions; Prolonged emotional arousal or rumination; Struggles with calming down after distress; Sensitivity to perceived rejection or invalidation; Impulsive behaviors in response to emotions; Emotional overwhelm in interpersonal settings; Guilt or shame following emotional outbursts \\
    \midrule
    Career & Uncertainty about career path or direction; Mismatch between values and work roles; Fear of failure or underperformance at work; Workplace dynamics and interpersonal challenges; Career transitions and instability; Imposter feelings in professional environments; Burnout and chronic work-related stress; Conflict between ambition and well-being \\
    \midrule
    Alcohol & Escalation of alcohol use in response to stress; Social pressure and drinking culture; Blackouts or memory impairment after drinking; Impact of alcohol on relationships or academics; Loss of control over drinking behavior; Guilt or shame following alcohol use; Withdrawal symptoms and physical consequences; Ambivalence about reducing alcohol consumption \\
    \midrule
    Perfectionism & Unrealistically high self-imposed standards; Fear of mistakes and failure; Chronic dissatisfaction despite success; Over-identification with achievement; Procrastination linked to performance anxiety; Rigidity in goal-setting and evaluation; Self-worth tied to productivity or outcomes; Interpersonal strain due to perfectionistic tendencies \\
    \midrule
    Harassment / Emotional Abuse & Verbal degradation and humiliation; Manipulation and gaslighting; Patterns of control or coercion; Emotional invalidation and dismissal; Fear and hypervigilance in relationships; Loss of autonomy or self-confidence; Isolation from social support systems; Difficulty recognizing or labeling abuse \\
    \midrule
    Drugs & Recreational use turning into dependency; Use of substances to escape emotions; Impact of drug use on functioning or goals; Risk-taking behaviors under influence; Tolerance, withdrawal, and physiological effects; Stigma and secrecy surrounding drug use; Interpersonal conflict over substance use; Cycles of use, regret, and attempted control \\
        \midrule
    Self-Injurious Thoughts or Behaviors & Urges to self-harm during emotional distress; Use of self-injury for emotion regulation; Secrecy, shame, or hiding of wounds; Triggers related to interpersonal conflict; Distinction between suicidal and non-suicidal intent; Cycles of guilt and relief after self-injury; Difficulty expressing pain verbally; Concern from others and responses to disclosure \\
    \midrule
    Health / Medical & Chronic illness and emotional adjustment; Uncertainty and fear about health outcomes; Medical trauma and negative healthcare experiences; Changes in self-identity due to illness; Impact of physical symptoms on daily life; Dependency on caregivers or medical systems; Health-related stigma or isolation; Balancing treatment adherence with quality of life \\
    \midrule
    Anger Management & Triggers and precipitating factors; Impulse control difficulties; Communication challenges during anger; Emotional awareness and regulation; Patterns of reactive versus suppressed anger; Impact of anger on relationships; Cognitive interpretations fueling anger; Cultural or familial attitudes toward anger \\
    \midrule
    Financial & Stress related to debt or financial insecurity; Pressure to support family or dependents; Impact of finances on academic or career decisions; Conflict over money in relationships; Financial instability during life transitions; Shame or stigma around financial struggles; Decision-making under financial stress; Access to basic needs and economic survival \\
    \midrule
    Physical Abuse / Assault & Fear of re-experiencing physical harm; Hypervigilance and startle responses; Control and power dynamics in abusive relationships; Shame, secrecy, and self-blame after assault; Avoidance of physical contact or intimacy; Medical consequences or visible injuries; Barriers to seeking help or reporting abuse; Rebuilding safety and bodily autonomy \\
    \midrule
    Mood Instability (Bipolar Symptoms) & Rapid shifts between emotional states; Elevated mood and high energy periods; Impulsivity and risk-taking during mood peaks; Low mood and fatigue during depressive episodes; Disturbed sleep and circadian disruptions; Impact of mood swings on relationships or work; Difficulties with emotional self-tracking; Fluctuating motivation and goal-directed behavior \\
    \midrule
    Obsessions or Compulsions & Intrusive, repetitive thoughts or fears; Rituals or behaviors used to reduce anxiety; Fear of contamination or harm; Mental checking and reassurance-seeking; Distress caused by loss of control over thoughts; Time-consuming routines impacting functioning; Shame or secrecy about compulsive behaviors; Difficulty distinguishing thoughts from reality \\
    \midrule
    Racial, Ethnic, or Cultural Concerns & Experiences of discrimination or bias; Cultural identity conflict or assimilation pressure; Stereotype threat and performance anxiety; Lack of representation or cultural understanding; Intergenerational cultural value differences; Microaggressions and subtle invalidation; Belongingness in predominantly different spaces; Navigating dual or multiple cultural identities \\
        \midrule
    Sexual Orientation & Confusion or questioning of sexual identity; Fear of rejection or discrimination; Coming out and identity disclosure challenges; Internalized stigma or self-judgment; Navigating intimacy and romantic relationships; Lack of familial or community support; Intersection with cultural or religious beliefs; Isolation or invisibility in social spaces \\
    \midrule
    Sexual Concern & Anxiety or distress related to sexual performance; Mismatched sexual desire in relationships; Body image concerns affecting sexual confidence; Trauma-related sexual avoidance or discomfort; Confusion around sexual norms or boundaries; Changes in libido due to stress or health issues; Moral or cultural conflict around sexuality; Feelings of guilt, shame, or inadequacy \\
    \midrule
    Legal / Judicial / Conduct & Involvement with law enforcement or legal proceedings; Anxiety about disciplinary action or consequences; Behavioral impulsivity leading to legal risk; Stigma and social judgment after legal issues; Family or peer conflict related to conduct concerns; Difficulties with authority or institutional systems; Reintegration after punishment or adjudication; Ethical conflict and personal accountability \\
    \midrule
    Religion / Spirituality & Spiritual identity exploration and development; Conflict between beliefs and lived experiences; Religious doubt or existential questioning; Feeling disconnected from spiritual practices; Experiences of religious guilt or shame; Spiritual coping during life stressors; Navigating faith in diverse or secular settings; Religious trauma or negative past experiences \\
    \midrule
    Gender Identity & Exploration and affirmation of gender identity; Gender dysphoria and body-related distress; Coming out and social transition challenges; Misgendering and invalidation experiences; Intersection of gender with culture or religion; Barriers to accessing gender-affirming care; Navigating relationships during gender transition; Safety concerns and discrimination risk \\
    \midrule
    Dissociative Experiences & Depersonalization or feeling disconnected from self; Derealization and distorted perception of reality; Memory gaps or time loss during daily activities; Emotional detachment and numbness; Altered sense of identity or fragmentation; Dissociation triggered by stress or trauma; Impact of dissociation on daily functioning; Fear or confusion around dissociative episodes \\
    \midrule
    Violent Thoughts or Behaviors Toward Others & Intrusive aggressive thoughts or fantasies; Impulsive outbursts during emotional distress; History of physical aggression or altercations; Anger misdirected toward others; Fear of harming someone unintentionally; Difficulty managing interpersonal conflict; Hostility related to perceived injustice; Impact of aggression on relationships and trust \\
    \midrule
    Autism Spectrum & Sensory sensitivities and overstimulation; Challenges with nonverbal communication; Difficulties in social reciprocity; Restricted or repetitive interests and behaviors; Need for structure and predictability; Masking and social camouflaging; Co-occurring anxiety or emotional regulation issues; Navigating relationships and neurotypical expectations \\
    \midrule
    Learning Disorder / Disability & Academic challenges despite effort; Difficulty with reading, writing, or math tasks; Processing speed or working memory difficulties; Self-esteem impacted by learning differences; Frustration with traditional learning environments; Disparities between potential and performance; Need for accommodations or support; Misunderstanding or stigma around learning differences \\
    \midrule
    Discrimination & Experiences of bias based on identity; Microaggressions in daily interactions; Internalized oppression or inferiority; Navigating exclusion or underrepresentation; Impact of systemic inequities on mental health; Fear of visibility or disclosure; Coping with invalidation or erasure; Social or institutional barriers to inclusion \\
    \midrule
    Psychotic Thoughts or Behaviors & Delusions or fixed false beliefs; Auditory or visual hallucinations; Paranoia and suspicious thinking; Disorganized thoughts or speech patterns; Emotional distress linked to altered perceptions; Fear of losing touch with reality; Social withdrawal due to perceptual changes; Uncertainty distinguishing internal vs external stimuli \\
    \midrule
    Addiction (Not Drugs or Alcohol) & Compulsive engagement in specific behaviors; Loss of control over time or frequency of use; Negative consequences of continued behavior; Emotional reliance on the addictive activity; Cravings or urges triggered by distress; Concealing behavior from others; Withdrawal symptoms when abstaining; Interference with goals or responsibilities \\
    \midrule
    Stalking & Persistent unwanted attention or contact; Fear or anxiety about personal safety; Monitoring or surveillance behavior by others; Loss of privacy and autonomy; Emotional toll of being followed or harassed; Legal or institutional response to stalking; Disruption of daily routines due to fear; Relationship between victim and perpetrator \\
    \midrule
    Pregnancy-Related & Emotional adjustment to pregnancy; Anxiety about childbirth and motherhood; Body image and physical changes during pregnancy; Pregnancy-related grief or loss (e.g., miscarriage); Conflict between pregnancy and life goals; Unplanned or ambivalent pregnancy feelings; Impact of pregnancy on relationships; Health complications and medical concerns \\

    \bottomrule
\end{longtable}

\newpage
\section{LLM Subtopic Categorization Prompt}

The following prompt was used to classify queries and documents into their corresponding main topics and subtopics.

\label{appendix:topic_prompt}
\begin{lstlisting}[
  breaklines=true,
  breakatwhitespace=true,
  columns=fullflexible,
  keepspaces=true,
  literate={-}{{-}}1 {–}{{-}}1 {—}{{-}}1
           {‘}{{'}}1 {’}{{'}}1
           {"}{{"}}1 {"}{{"}}1
           {±}{{$\pm$}}1
]
You are a mental health assistant. Given a piece of text, select up to **three** subtopics
from the list that are most relevant to the content. These subtopics will help with
**categorizing the text for mental health document retrieval**.

Pick subtopics that are **clearly reflected**, **implicitly referenced**, or **closely related**
to the text.

Then, return a **weighted relevance distribution** across those selected subtopics. The weights
should reflect how strongly each subtopic is present in the text, and they must **sum to 1.0**.

### Subtopic Scoring Rubric:
Use the following guidelines to assign weights:

- **0.6 – 0.8**  The subtopic is the **main focus** (central concern, emphasized throughout, or explicitly repeated)
- **0.2 – 0.4**  The subtopic is a **strong secondary theme** (clearly relevant but not dominant)
- **0.05 – 0.2** The subtopic is **briefly or implicitly mentioned**
- **0.0**        The subtopic is **not relevant** — do **not** include it

Only include **up to three subtopics** with nonzero weights. Use only the subtopics from the list below.
Do **not** create new subtopics.

### Output Format:
1. A **JSON object** with subtopic names as keys and float weights as values
2. A separate line stating the **primary subtopic** — the one with the highest weight

### Available Subtopics:
{subtopics}

### Text:
{text}

### Example Output:
{
  "ADHD: Neurodiversity": 0.7,
  "Reparenting your inner child: Inner child and healing": 0.2,
  "Coping with guilt: Shame and guilt": 0.1
}
\end{lstlisting}

\section{Ranked Subtopics by Hybrid Gap Score}
\label{appendix:gap_scores}

Table~\ref{tab:ranked_subtopics} lists subtopics with their corresponding main topics, coverage scores, usefulness scores, and combined hybrid gap scores.
Higher scores (between 1--100) indicate larger content gaps and thus a greater need to add documents.
If the usefulness score is marked as \textit{N/A}, this means that no documents were available for that subtopic.
If the coverage score is low but the usefulness score is high (e.g., career transitions and instability), it indicates that although few resources exist on this topic, those available are highly relevant and valuable to user queries. The hybrid score was calculated as a 50\% coverage score and 50\% usefulness score when both were available, or as 100\% coverage when usefulness was \textit{N/A}.



{\scriptsize
\begin{longtable}{p{0.22\linewidth} p{0.34\linewidth} p{0.12\linewidth} p{0.12\linewidth} p{0.12\linewidth}}
\caption{Main topics, subtopics, and gap scores.\label{tab:ranked_subtopics}}\\

\toprule
\textbf{Main Topic} & \textbf{Subtopic} & \textbf{Coverage Score (1--100)} &
\textbf{Usefulness Score (1--100)} & \textbf{Hybrid Score (1--100)} \\
\midrule
\endfirsthead

\toprule
\textbf{Main Topic} & \textbf{Subtopic} & \textbf{Coverage Score (1--100)} &
\textbf{Usefulness Score (1--100)} & \textbf{Hybrid Score (1--100)} \\
\midrule
\endhead

\midrule
\multicolumn{5}{r}{\textit{Continued on next page}} \\
\midrule
\endfoot

\bottomrule
\endlastfoot

    Depression & Self-critical thoughts and low self-worth & 100 & N/A & \textbf{100} \\
    Relationship Problems & Trust erosion and boundary issues & 82.04 & 62 & \textbf{72.02} \\
    Anxiety & Fear of illness and health-related vigilance & 75.80 & 65 & \textbf{70.40} \\
    Relationship Problems & Attachment insecurity and emotional distance & 73.44 & 62 & \textbf{67.72} \\
    Emotional Dysregulation & Rapid mood fluctuations and reactivity & 61.38 & 60 & \textbf{60.69} \\
    Family & Parental conflict and household tension & 58.98 & 60 & \textbf{59.49} \\
    Depression & Social isolation and disconnection & 55.42 & 60 & \textbf{57.71} \\
    Depression & Anhedonia and withdrawal from rewarding activities & 54.72 & 60 & \textbf{57.36} \\
    Anxiety & Social evaluation concerns and avoidance & 56.72 & 56 & \textbf{56.36} \\
    Anxiety & Sleep disturbances linked to anxiety & 55.48 & 57 & \textbf{56.24} \\
    Sexual Orientation & Navigating intimacy and romantic relationships & 58.50 & 50 & \textbf{54.25} \\
    Addiction (Not Drugs or Alcohol) & Other & 54.25 & N/A & \textbf{54.25} \\
    Interpersonal Functioning & People-pleasing and fear of rejection & 22.76 & 82.54 & \textbf{52.65} \\
    Career & Career transitions and instability & 19.77 & 84.36 & \textbf{52.06} \\
    Attention / Concentration Difficulties & Other & 28.13 & 75.54 & \textbf{51.84} \\
    Career & Fear of failure or underperformance at work & 50.79 & N/A & \textbf{50.79} \\
    Relationship Problems & Romantic conflict and unresolved tension & 54.12 & 45.49 & \textbf{49.80} \\
    Relationship Problems & Navigating long-distance or commitment uncertainty & 24.94 & 74.07 & \textbf{49.51} \\
    Suicidality & Barriers to help-seeking or disclosure & 32.24 & 62.95 & \textbf{47.60} \\
    Career & Uncertainty about career path or direction & 21.73 & 73.45 & \textbf{47.59} \\
    Interpersonal Functioning & Patterns of miscommunication or avoidance & 22.29 & 72.40 & \textbf{47.34} \\
    Addiction (Not Drugs or Alcohol) & Compulsive engagement in specific behaviors & 25.60 & 67.57 & \textbf{46.58} \\
    Addiction (Not Drugs or Alcohol) & Emotional reliance on the addictive activity & 12.47 & 78.69 & \textbf{45.58} \\
    Adjustment to New Environment & Loss of familiarity and support systems & 22.76 & 67.85 & \textbf{45.31} \\
    Relationship Problems & Breakups, separation, and emotional recovery & 47.09 & 43.42 & \textbf{45.25} \\
    Identity Development & Internal conflict around authenticity & 14.06 & 75.54 & \textbf{44.80} \\
    Self-esteem / Confidence & Fear of judgment and social comparison & 36.71 & 51.47 & \textbf{44.09} \\
    Anxiety & Impact of anxiety on daily functioning & 37.79 & 50.25 & \textbf{44.02} \\
    Anxiety & Cognitive rumination and overthinking loops & 51.20 & 36.84 & \textbf{44.02} \\
    Anger Management & Impulse control difficulties & 24.94 & 62.95 & \textbf{43.95} \\
    Social Isolation & Emotional withdrawal from connection & 43.76 & N/A & \textbf{43.76} \\
    Eating / Body Image & Emotional or binge eating behaviors & 43.76 & N/A & \textbf{43.76} \\
    Harassment / Emotional Abuse & Manipulation and gaslighting & 43.76 & N/A & \textbf{43.76} \\
    Dissociative Experiences & Emotional detachment and numbness & 43.76 & N/A & \textbf{43.76} \\
    Self-esteem / Confidence & Confidence in academic or work performance & 11.38 & 75.54 & \textbf{43.46} \\
    Family & Communication breakdowns within the family & 31.95 & 54.91 & \textbf{43.43} \\
    Depression & Feelings of hopelessness or helplessness & 55.91 & 30.27 & \textbf{43.09} \\
    Stress & Interpersonal sources of stress & 27.30 & 57.50 & \textbf{42.40} \\
    Depression & Emotional flatness and loss of affect & 60.59 & 23.52 & \textbf{42.05} \\
    Anxiety & Panic symptoms and physiological reactivity & 57.90 & 25.41 & \textbf{41.66} \\
    Academic Performance & Procrastination and avoidance cycles & 44.58 & 38.56 & \textbf{41.57} \\
    Racial, Ethnic, or Cultural Concerns & Experiences of discrimination or bias & 9.90 & 71.35 & \textbf{40.62} \\
    Stress & Work-life balance and professional burnout & 37.79 & 41.50 & \textbf{39.64} \\
    Family & Parent-child relationship challenges & 34.72 & 44.33 & \textbf{39.52} \\
    Grief / Loss & Emptiness and identity confusion & 39.31 & N/A & \textbf{39.31} \\
    Academic Performance & Difficulty with executive functioning & 39.31 & N/A & \textbf{39.31} \\
    Family & Family roles and expectations & 39.31 & N/A & \textbf{39.31} \\
    Career & Burnout and chronic work-related stress & 39.31 & N/A & \textbf{39.31} \\
    Self-esteem / Confidence & Perfectionism and fear of failure & 32.66 & 44.07 & \textbf{38.36} \\
    Grief / Loss & Shock and disbelief following loss & 36.36 & 40.29 & \textbf{38.32} \\
    Interpersonal Functioning & Assertiveness and self-expression difficulties & 27.30 & 48.26 & \textbf{37.78} \\
    Health / Medical & Chronic illness and emotional adjustment & 19.77 & 55.08 & \textbf{37.43} \\
    Self-esteem / Confidence & Negative self-concept and inner critic & 34.15 & 40.17 & \textbf{37.16} \\
    Health / Medical & Medical trauma and negative healthcare experiences & 22.29 & 50.36 & \textbf{36.33} \\
    Family & Navigating major family transitions or disruptions & 10.56 & 60.85 & \textbf{35.71} \\
    Suicidality & Suicidal ideation during emotional distress & 40.20 & 29.06 & \textbf{34.63} \\
    Harassment / Emotional Abuse & Patterns of control or coercion & 16.74 & 52.46 & \textbf{34.60} \\
    Anxiety & Uncertainty intolerance and future-focused fear & 28.96 & 39.35 & \textbf{34.15} \\
    Psychotic Thoughts or Behaviors & Auditory or visual hallucinations & 33.86 & N/A & \textbf{33.86} \\
    Autism Spectrum & Sensory sensitivities and overstimulation & 33.86 & N/A & \textbf{33.86} \\
    Financial & Financial instability during life transitions & 33.86 & N/A & \textbf{33.86} \\
    Academic Performance & Low academic motivation or engagement & 37.85 & 29.00 & \textbf{33.43} \\
    Social Isolation & Loneliness despite social presence & 22.29 & 44.07 & \textbf{33.18} \\
    Eating / Body Image & Preoccupation with weight and shape & 22.29 & 44.07 & \textbf{33.18} \\
    Social Isolation & Challenges in initiating or sustaining friendships & 22.76 & 43.37 & \textbf{33.07} \\
    Relationship Problems & Communication mismatches and misunderstandings & 33.18 & 32.26 & \textbf{32.72} \\
    Legal / Judicial / Conduct & Stigma and social judgment after legal issues & 14.06 & 50.36 & \textbf{32.21} \\
    Trauma & Avoidance of trauma reminders & 12.47 & 50.36 & \textbf{31.42} \\
    Alcohol & Escalation of alcohol use in response to stress & 22.29 & 39.66 & \textbf{30.98} \\
    Stress & Acute stress responses and short-term overwhelm & 33.47 & 28.33 & \textbf{30.90} \\
    Sleep & Sleep and daytime fatigue or functioning & 19.77 & 40.92 & \textbf{30.34} \\
    Stress & Academic and performance-related stress & 35.58 & 24.62 & \textbf{30.10} \\
    Harassment / Emotional Abuse & Verbal degradation and humiliation & 24.52 & 35.67 & \textbf{30.10} \\
    Pregnancy-related & Emotional adjustment to pregnancy & 19.77 & 40.29 & \textbf{30.03} \\
    Financial & Stress related to debt or financial insecurity & 23.84 & 35.53 & \textbf{29.69} \\
    Stress & Chronic stress and allostatic load & 33.15 & 24.67 & \textbf{28.91} \\
    Anxiety & Generalized worry and anticipatory tension & 35.48 & 21.54 & \textbf{28.51} \\
    Suicidality & Hopelessness and desire for relief & 46.72 & 9.79 & \textbf{28.26} \\
    Stress & Cognitive overload and decision fatigue & 20.65 & 35.36 & \textbf{28.01} \\
    Trauma & Hyperarousal and physiological reactivity & 22.29 & 31.48 & \textbf{26.88} \\
    Trauma & Emotional numbing and detachment & 22.29 & 31.48 & \textbf{26.88} \\
    Self-injurious Thoughts or Behaviors & Urges to self-harm during emotional distress & 26.83 & N/A & \textbf{26.83} \\
    Other & Stalking & 26.83 & N/A & \textbf{26.83} \\
    Obsessions or Compulsions & Other & 26.83 & N/A & \textbf{26.83} \\
    Self-esteem / Confidence & Fear of rejection and social comparison & 26.83 & N/A & \textbf{26.83} \\
    Violent Thoughts or Behaviors Toward Others & Intrusive aggressive thoughts or fantasies & 26.83 & N/A & \textbf{26.83} \\
    Emotional Dysregulation & Impulsive behaviors in response to emotions & 26.83 & N/A & \textbf{26.83} \\
    Gender Identity & Exploration and affirmation of gender identity & 26.83 & N/A & \textbf{26.83} \\
    Emotional Dysregulation & Emotional overwhelm in interpersonal settings & 26.83 & N/A & \textbf{26.83} \\
    Self-esteem / Confidence & Imposter feelings and perceived inadequacy & 26.83 & N/A & \textbf{26.83} \\
    Dissociative Experiences & Depersonalization or feeling disconnected from self & 26.83 & N/A & \textbf{26.83} \\
    Interpersonal Functioning & Conflict resolution and emotional regulation & 24.52 & 27.59 & \textbf{26.06} \\
    Harassment / Emotional Abuse & Emotional invalidation and dismissal & 14.06 & 37.77 & \textbf{25.92} \\
    Self-esteem / Confidence & Body image and physical self-perception & 19.77 & 31.48 & \textbf{25.62} \\
    Eating / Body Image & Restrictive eating patterns and control & 22.76 & 27.98 & \textbf{25.37} \\
    Adjustment to New Environment & Uncertainty and identity shifts & 11.38 & 37.77 & \textbf{24.58} \\
    Self-injurious Thoughts or Behaviors & Use of self-injury for emotion regulation & 11.38 & 37.77 & \textbf{24.58} \\
    Eating / Body Image & Body dissatisfaction and self-surveillance & 27.80 & 20.70 & \textbf{24.25} \\
    Suicidality & Passive versus active suicidal thoughts & 22.29 & 0 & \textbf{22.29} \\
    Sexual Abuse / Assault & Shame and internalized self-blame & 22.29 & 0 & \textbf{22.29} \\
    Depression & Cognitive distortions and negative thought patterns & 22.85 & 20.49 & \textbf{21.67} \\
    Academic Performance & Test anxiety and performance pressure & 19.77 & 22.03 & \textbf{20.90} \\
    Identity Development & Uncertainty in self-definition or direction & 19.77 & 19.83 & \textbf{19.80} \\
    Trauma & Intrusive memories and flashbacks & 22.76 & 16.79 & \textbf{19.78} \\
    Self-injurious Thoughts or Behaviors & Difficulty expressing pain verbally & 16.93 & N/A & \textbf{16.93} \\
    Autism Spectrum & Navigating relationships and neurotypical expectations & 16.93 & N/A & \textbf{16.93} \\
    Emotional Dysregulation & Sensitivity to perceived rejection or invalidation & 16.93 & N/A & \textbf{16.93} \\
    Mood Instability (Bipolar Symptoms) & Fluctuating motivation and goal-directed behavior & 16.93 & N/A & \textbf{16.93} \\
    Anger Management & Impact of anger on relationships & 16.93 & N/A & \textbf{16.93} \\
    Attention / Concentration Difficulties & Task initiation and follow-through issues & 16.93 & N/A & \textbf{16.93} \\
    Self-injurious Thoughts or Behaviors & Relapse on cutting and self-injury & 16.93 & N/A & \textbf{16.93} \\
    Career & Workplace dynamics and interpersonal challenges & 16.93 & N/A & \textbf{16.93} \\
    Health / Medical & Uncertainty and fear about health outcomes & 16.93 & N/A & \textbf{16.93} \\
    Stress & Stress-related physical symptoms & 16.93 & N/A & \textbf{16.93} \\
    Sexual Concern & Trauma-related sexual avoidance or discomfort & 16.93 & N/A & \textbf{16.93} \\
    Sleep & Sleep maintenance and frequent awakenings & 16.93 & N/A & \textbf{16.93} \\
    Anger Management & Triggers and precipitating factors & 16.93 & N/A & \textbf{16.93} \\
    Trauma & Shame, guilt, or self-blame post-trauma & 16.93 & N/A & \textbf{16.93} \\
    Mood Instability (Bipolar Symptoms) & Fear of mental health conditions & 16.93 & N/A & \textbf{16.93} \\
    Obsessions or Compulsions & Rituals or behaviors used to reduce anxiety & 16.93 & N/A & \textbf{16.93} \\
    Career & Imposter feelings in professional environments & 16.93 & N/A & \textbf{16.93} \\
    Emotional Dysregulation & Emotional dysregulation & 16.93 & N/A & \textbf{16.93} \\
    Relationship Problems & Jealousy and insecurity in relationships & 16.93 & N/A & \textbf{16.93} \\
    Academic Performance & Fear of failure and perfectionistic standards & 16.93 & N/A & \textbf{16.93} \\
    Legal / Judicial / Conduct & Anxiety about disciplinary action or consequences & 16.93 & N/A & \textbf{16.93} \\
    Mood Instability (Bipolar Symptoms) & Emotional dysregulation and mood instability & 16.93 & N/A & \textbf{16.93} \\
    Emotional Dysregulation & Other & 16.93 & N/A & \textbf{16.93} \\
    Emotional Dysregulation & Emotion regulation struggles and emotional reactivity & 16.93 & N/A & \textbf{16.93} \\
    Social Isolation & Disconnection from meaningful relationships & 16.93 & N/A & \textbf{16.93} \\
    Grief / Loss & Disruption to roles and daily routines & 16.93 & N/A & \textbf{16.93} \\
    Emotional Dysregulation & Emotional dysregulation & 16.93 & N/A & \textbf{16.93} \\
    Psychotic Thoughts or Behaviors & Delusions or fixed false beliefs & 16.93 & N/A & \textbf{16.93} \\
    Perfectionism & Comparative social concern and self-worth & 16.93 & N/A & \textbf{16.93} \\
    Sexual Orientation & Confusion or questioning of sexual identity & 16.93 & N/A & \textbf{16.93} \\
    Depression & Emotional flatness and loss of affect & 16.93 & N/A & \textbf{16.93} \\
    Dissociative Experiences & Dissociation triggered by stress or trauma & 16.93 & N/A & \textbf{16.93} \\
    Trauma & Disrupted memory or time perception & 16.93 & N/A & \textbf{16.93} \\
    Dissociative Experiences & Derealization and distorted perception of reality & 16.93 & N/A & \textbf{16.93} \\
    Academic Performance & Fear of falling behind & 16.93 & N/A & \textbf{16.93} \\
    Autism Spectrum & Navigation of long-term relationships and deep connection & 16.93 & N/A & \textbf{16.93} \\
    Psychotic Thoughts or Behaviors & Other & 16.93 & N/A & \textbf{16.93} \\
    Relationship Problems & Impact of external stressors on relationships & 16.93 & N/A & \textbf{16.93} \\
    Suicidality & Impact of suicidality on relationships & 16.93 & N/A & \textbf{16.93} \\
    Health / Medical & Impact of physical symptoms on daily life & 16.93 & N/A & \textbf{16.93} \\
    Sleep & Insomnia and difficulties initiating sleep & 16.93 & N/A & \textbf{16.93} \\
    Depression & Major depression and relational dynamics & 16.93 & N/A & \textbf{16.93} \\
    Anger Management & Patterns of reactive versus suppressed anger & 16.93 & N/A & \textbf{16.93} \\
    Sexual Concern & Mismatched sexual desire in relationships & 16.93 & N/A & \textbf{16.93} \\
    Sexual Abuse / Assault & Fear and hypervigilance in safe spaces & 16.93 & N/A & \textbf{16.93} \\
    Emotional Dysregulation & Emotional dysregulation and impulsivity & 16.93 & N/A & \textbf{16.93} \\
    Emotional Dysregulation & Emotional dysregulation and difficulty managing emotions & 16.93 & N/A & \textbf{16.93} \\
    Emotional Dysregulation & Difficulty identifying or naming emotions & 16.93 & N/A & \textbf{16.93} \\
    Anxiety & Jealousy and insecurity in relationships & 16.93 & N/A & \textbf{16.93} \\
    Pregnancy-related & Postpartum depression and mood changes & 16.93 & N/A & \textbf{16.93} \\
    Relationship Problems & Jealousy in friendships & 16.93 & N/A & \textbf{16.93} \\
    Perfectionism & Interpersonal strain due to perfectionistic tendencies & 16.93 & N/A & \textbf{16.93} \\
    Mood Instability (Bipolar Symptoms) & Impulsivity and risk-taking during mood peaks & 16.93 & N/A & \textbf{16.93} \\
    Emotional Dysregulation & Prolonged emotional arousal or rumination & 16.93 & N/A & \textbf{16.93} \\
    Relationship Problems & Breakups, separation, and emotional recovery & 16.93 & N/A & \textbf{16.93} \\
    Mood Instability (Bipolar Symptoms) & Rapid shifts between emotional states & 16.93 & N/A & \textbf{16.93} \\
    Self-injurious Thoughts or Behaviors & Use of self-injury for emotion regulation & 16.93 & N/A & \textbf{16.93} \\
    Grief / Loss & Other & 16.93 & N/A & \textbf{16.93} \\
    Adjustment to New Environment & Isolation during transitional periods & 16.93 & N/A & \textbf{16.93} \\
    Psychotic Thoughts or Behaviors & Paranoia and suspicious thinking & 16.93 & N/A & \textbf{16.93} \\
    Grief / Loss & Sadness and yearning for what was lost & 18.72 & 13.99 & \textbf{16.35} \\
    Sexual Abuse / Assault & Violation of bodily autonomy and trust & 16.74 & 14.69 & \textbf{15.71} \\
    Interpersonal Functioning & Boundary setting and personal space & 10.56 & 16.79 & \textbf{13.67} \\
    Social Isolation & Perceived rejection or exclusion & 14.06 & 12.59 & \textbf{13.33} \\
    Self-esteem / Confidence & Self-validation versus external validation & 8.90 & 16.79 & \textbf{12.84} \\
    Obsessions or Compulsions & Intrusive, repetitive thoughts or fears & 17.79 & 7.83 & \textbf{12.81} \\
    Stress & Coping capacity and resilience thresholds & 12.56 & 11.16 & \textbf{11.86} \\
    
\end{longtable}
}

\section{Therapist-Reviewed LLM Usefulness Scoring Prompt for Document Helpfulness}

To ground our evaluation in clinical relevance, therapists designed the rubric to measure how effectively a document responds to a given user query.

\label{appendix:rubric}
\begin{lstlisting}[
  breaklines=true,
  breakatwhitespace=true,
  columns=fullflexible,
  keepspaces=true,
  literate={-}{{-}}1 {–}{{-}}1 {—}{{-}}1
           {‘}{{'}}1 {’}{{'}}1
           {"}{{"}}1 {"}{{"}}1
           {±}{{$\pm$}}1
]
# Role: Therapist Evaluating Document Relevance

## Goal
You are a licensed mental health professional reviewing documents retrieved in response to user-submitted queries. These users are actively seeking to improve their mental health. Your task is to evaluate how appropriate and effective each document is in addressing a specific user concern.

You will be shown:
* A user’s query or concern
* A document retrieved from a mental health resource corpus

You must assign a **total Relevance Score from 1 to 100**, where:
* **Contextual Relevance** contributes **1-50** points
* **Practical Helpfulness & Engagement** contributes **1-50** points

The final score should be the **sum of both sub-scores**.

---

## Criterion 1: Contextual Relevance (1-50)
**Definition:** How well the recommendation matches the user’s stated needs, concerns, and emotional state.

**Evaluator Prompts:**
* Does the resource address the core concern expressed by the user?
* Is the emotional tone appropriate for their state (e.g., gentle for distress, motivational for low energy)?
* Does it address specific aspects of the concern rather than being overly generic?
* Is it timely for their situation (immediate relief vs. long-term growth)?
* Does it demonstrate an understanding of the user’s context (age, culture, life situation)?

**Range Interpretation**
* **1–10:** Off-topic, unrelated, or mismatches the concern entirely.
* **11–20:** Slightly touches on the concern but is mostly generic or irrelevant.
* **21–30:** Somewhat relevant—covers parts of the concern but misses nuance or key details.
* **31–40:** Mostly relevant—well aligned with the concern, minor gaps.
* **41–50:** Highly relevant—directly and deeply addresses the concern with excellent contextual fit.

---

## Criterion 2: Practical Helpfulness & Engagement (1-50)
**Definition:** How helpful, easy, motivating, and realistic it is for the user to follow and benefit from the recommendation.

**Evaluator Prompts:**
* Is the content easy to understand (clear language, minimal jargon)?
* Is it practical for the user’s likely constraints (time, cost, device, location)?
* Does the format fit the user’s likely engagement style (article, video, exercise, interactive tool)?
* Is it motivating or encouraging enough to prompt action?
* Does it include clear next steps or pathways for ongoing support?
* Does it feel doable in the user’s daily life without overwhelming them?

**Range Interpretation**
* **1–10:** Offers no actionable help or emotional value; impractical or inaccessible.
* **11–20:** Provides minimal insight or comfort; hard to engage with or unrealistic for the user.
* **21–30:** Some useful guidance or support but notable barriers (clarity, format, feasibility).
* **31–40:** Solid practical help—clear, motivating, and mostly easy to follow.
* **41–50:** Highly helpful—validating, empowering, practical, and easy to integrate into daily life.

---

## Scoring Bands (for overall 1–100 score; informational only)
* **90–100:** Excellent — highly relevant, practical, and engaging.
* **70–89:** Good — meets most needs, with minor gaps.
* **50–69:** Moderate — some relevance or usability, but notable limitations.
* **<50:** Poor — unlikely to be helpful or actionable.

---

## Output Instructions
* Assign scores out of 50 for each criterion and **sum them** to produce a **final score between 2 and 100**.
* **Do NOT include your reasoning.**
* **Respond ONLY with a single integer between 1 and 100. No extra text, decimals, or formatting.**

---

## Input
**User Query:**
{user_query}

**Retrieved Document:**
{retrieved_document}

---

## Output Format
Relevance Score (1–100): <insert score here>
\end{lstlisting}

\section{Impact of Coverage-Usefulness Weighting on Corpus Allocation}

\begin{table}[h]
  \caption{Impact of different coverage/usefulness weight combinations on targeted corpus allocation changes relative to a 50/50 baseline.}
  \label{appendix:coverage_usefulness_weights}
  \label{tab:hybrid_weights}
  \centering
  \scalebox{0.86}{
    \begin{tabular}{ccc}
      \toprule
      Weight Combination (Coverage/Usefulness) & Avg. Absolute Difference (docs/subtopic) & Corpus \% Diff. \\
      \midrule
      40/60   & 0.205  & 3.5  \\
      60/40   & 0.263  & 4.5  \\
      30/70   & 0.474  & 8.1  \\
      70/30   & 0.497  & 8.5  \\
      100/0   & 1.264  & 21.6 \\
      0/100   & 1.415  & 24.2 \\
      \bottomrule
    \end{tabular}
  }
\end{table}

\section{LLM Prompt for Synthetic Document Generation}

The following prompt was used to generate synthetic documents derived from the publicly available metadata.

\label{appendix:synthetic_prompts}
\begin{lstlisting}[
  breaklines=true,
  breakatwhitespace=true,
  columns=fullflexible,
  keepspaces=true,
  literate={-}{{-}}1 {–}{{-}}1 {—}{{-}}1
           {‘}{{'}}1 {’}{{'}}1
           {"}{{"}}1 {"}{{"}}1
           {±}{{$\pm$}}1
]
# Role: System (Synthetic Document Generator)

You are playing the role of an **expert human writer**. Your task is to generate a **realistic, high-quality, and emotionally resonant article** based on the provided metadata. The result should be entirely original and plausible, as if written for an actual audience on a mental health or wellbeing platform.

------------------------------------------------------------
## Therapist-Informed Writing Style
- When crafting the article, **refer implicitly to the tone, values, and clarity standards observed in therapist-written materials**.
- Avoid over-generalization or unrealistic optimism. Aim to reflect the kind of **supportive, balanced, and user-sensitive language** that a therapist might use.
- Prioritize **emotional safety, inclusivity, validation, and harm reduction**.

------------------------------------------------------------
## Core Guidelines for Article Generation

1. **Metadata as Inspiration**
   - Do **not** use the provided title, subtitle, or section headers directly.
   - Instead, **rewrite a new title, subtitle, and section headers** based on the underlying topic and intent.
   - Ensure the rewritten elements reflect the **core theme** and **emotional purpose**.

2. **Audience Inference**
   - Infer the likely **target audience** (e.g., teens, adults in crisis, caregivers, wellness-seekers).
   - Write with that audience’s **emotional state, context, and needs** in mind.
   - Use inclusive, clear language suitable for a **general or moderately informed audience**.

3. **Tone Emulation**
   - Determine the appropriate **tone** from the metadata (especially the rewritten subtitle) and maintain it throughout.
   - Avoid clinical or overly academic styles; aim for a **natural, emotionally supportive, lightly narrative voice**.

4. **Goal Alignment**
   - Infer the article’s **purpose** (e.g., help someone feel less alone; offer encouragement; provide coping strategies; encourage values or connection).
   - Let that goal **shape the structure and messaging**.

5. **Structure and Content**
   - **Title**: Rewrite an emotionally attuned title inspired by the metadata.
   - **Subtitle**: Write a thoughtful introductory subtitle that sets tone.
   - **Sections**:
     - Rewrite new top-level section headers.
     - For each header, write one coherent and engaging section.
     - If any section includes **subheaders**, treat them as subsections with their own paragraphs or bullets.
     - Vary sentence structure and use meaningful transitions.
   - **Length**: Aim for a total word count close to 'word_count' (±10%).

6. **Writing Quality**
   - Produce fluid, high-quality prose with varied sentence length and natural phrasing.
   - Maintain factual plausibility and psychological/emotional realism.
   - The article should feel like a reputable site’s content — but must be **completely original**.

7. **Ethical and Legal Compliance**
   - Do not copy from or reference any real-world source.
   - Avoid direct quotes or close paraphrases of existing articles.
   - Do not mention real names, authors, organizations, or events.

8. **Output Constraints**
   - Do **not** output the metadata or repeat the instructions.
   - Output only the article text (title, subtitle, sections).

------------------------------------------------------------
## Metadata (Use for structure, tone, audience, and goal inference):

{metadata}

## Begin Article Below:
\end{lstlisting}

\clearpage

\section{Comparative Analysis of Usefulness Scores Between Directed vs. Non-Directed Corpus Configurations Across RAG Pipelines}

\begin{table}[h]
  \small
  \caption{Average LLM usefulness scores (1--100) across four RAG pipelines for each \textbf{Directed} corpus configuration.}
   \label{tab:appendix_directed_rag_scores}
  \centering
  \scalebox{0.95}{
    \begin{tabular}{lccccc}
      \toprule
      & & \multicolumn{4}{c}{Retrieval Method} \\
      \cmidrule(r){3-6}
      Corpus & \% Increase from \textit{mindline.sg} & Baseline & Hierarchical & Reranking & Query Transformation \\
      \midrule
      \textit{\textit{mindline.sg}}  & 0\% (baseline)   & 54.57 & 65.12 & 68.44 & 66.97 \\
      Directed 1  & +12.9\%   & 56.44 & 70.84 & 74.01 & 75.76 \\
      Directed 2 & +41.9\%   & 57.87 & 74.78 & 76.87 & 78.86 \\
      Directed 3 & +74.4\%   & 58.92 & 77.20 & 78.63 & 80.12 \\
      Directed 4 & +129.2\%  & 60.16 & 78.21 & 80.78 & 80.46 \\
      Directed 5 & +232.0\%  & 61.85 & 79.86 & 80.78 & 81.41 \\
      Directed 6 & +317.8\%  & 62.90 & 79.76 & 81.33 & 81.58 \\
      Directed 7 & +403.1\%  & 63.33 & 80.08 & 81.55 & 81.37 \\
      Directed 8 & +542.1\%  & 64.01 & 80.36 & 81.54 & 81.47 \\
      Directed 9 & +661.8\%  & 64.40 & 80.09 & 81.59 & 81.62 \\
      Directed 10 & +763.3\%  & 64.68 & 80.29 & 81.48 & 81.96 \\
      Reference   & +1,974.2\% & 65.86 & 80.70 & 82.12 & 82.22 \\
      \bottomrule
    \end{tabular}
  }
\end{table}

\begin{table}[h]
  \small
  \caption{Average LLM usefulness scores (1--100) across four RAG pipelines for each non-\textbf{Non-Directed} corpus configuration.}
   \label{tab:appendix_nondirected_rag_scores}
  \centering
  \scalebox{0.95}{
  \begin{tabular}{lccccc}
    \toprule
    & & \multicolumn{4}{c}{Retrieval Method} \\
    \cmidrule(r){3-6}
    Corpus & \% Increase from \textit{mindline.sg} & Baseline & Hierarchical & Reranking & Query Transformation \\
    \midrule
    \textit{\textit{mindline.sg}}        & 0\% (baseline)   & 54.57 & 65.12 & 68.44 & 66.97 \\
    Non-Directed 1    & +12.9\%   & 55.22 & 66.63 & 69.54 & 68.81 \\
    Non-Directed 2   & +41.9\%   & 55.91 & 69.12 & 71.97 & 68.81 \\
    Non-Directed 3   & +74.4\%   & 56.80 & 71.23 & 73.72 & 74.64 \\
    Non-Directed 4   & +129.2\%  & 58.13 & 72.99 & 75.67 & 76.73 \\
    Non-Directed 5   & +232.0\%  & 59.64 & 75.08 & 76.88 & 78.17 \\
    Non-Directed 6   & +317.8\%  & 60.33 & 75.87 & 78.19 & 78.17 \\
    Non-Directed 7   & +403.1\%  & 60.63 & 76.99 & 78.90 & 79.61 \\
    Non-Directed 8   & +542.1\%  & 61.79 & 78.28 & 79.88 & 80.77 \\
    Non-Directed 9   & +661.8\%  & 62.14 & 78.65 & 80.46 & 80.42 \\
    Non-Directed 10   & +763.3\%  & 62.54 & 78.62 & 80.44 & 80.58 \\
    Reference         & +1,974.2\% & 65.86 & 80.70 & 82.12 & 82.22 \\
    \bottomrule
  \end{tabular}}
\end{table}

\end{document}